# Multiphasic pH profiles for the reaction of tris-(hydroxymethyl)-aminomethane with phenyl esters

Per Nissen

Norwegian University of Life Sciences
Department of Ecology and Natural Resource Management
P. O. Box 5003, NO-1432 Ås, Norway

per.nissen@nmbu.no

## Abstract

Reanalysis of data (Bruice and York 1961) for the pH dependences of the calculated apparent second-order rate constants ($k_2$') for the reaction of tris-(hydroxymethyl)-aminomethane (TRIS) with phenyl esters reveals that the pH profiles are consistently much better represented as multiphasic, i.e. as a series of straight lines separated by discontinuities in the form of sharp breaks or noncontiguities (jumps), than by the continuous function used by the authors. For *p*-nitrophenyl acetate, *m*-nitrophenyl acetate, *p*-chlorophenyl acetate and phenyl acetate, the profiles are best represented as multiphasic in plots of log $k_2$' vs. pH, whereas the profile for *p*-methylphenyl acetate is only multiphasic in a plot of $k_2$' vs. pH. Data for the pseudo-first-order rate constants ($k_{obs}$) give multiphasic pH profiles with patterns identical or very similar to those for $k_2$'. Extensive reanalyses of data for a wide variety of enzymatic and non-enzymatic reactions show, similarly, that pH profiles are in general better represented as multiphasic than as curvilinear.

## Introduction

Since the initial work of Michaelis and Davidsohn more than 100 years ago (Michaelis and Davidsohn 1911), pH dependences of the activities and kinetic parameters of enzymes have been usually, if not exclusively, represented by various forms of curvilinear profiles, often in the form of half-bells or bells, reflecting the assumed protonation/deprotonation of groups in the active site (or in the substrate). pH profiles for non-enzymatic reactions are usually also represented as curvilinear. Over the years, a large theoretical edifice has been constructed to help obtain information about mechanisms of enzyme action from the determination of pH dependences for enzymatic reactions as well as for reactions facilitated by small catalytic molecules in model systems. However, despite the apparent successes, the representation of pH profiles in the classical way may, nevertheless, be fundamentally in error. Given sufficiently detailed and precise data, most if not all pH profiles are better represented as multiphasic than as curvilinear. This will be exemplified in the following by reanalysis of data of Bruice and York (1961) for the reaction of TRIS with certain phenyl esters.

## Reanalysis

Data for $k_2$' and $k_{obs}$ were obtained from the Library of Congress (cf. note 14 in Bruice and York 1961). The authors' representation of the pH profile for log $k_2$' (thin lines in Fig. 1) were obtained from their equation 4 ($k_2' = k_b K_a/(K_a + \alpha_H) + k_a K_w K_a/(K_a + \alpha_H)$) and the values for $k_a$ and $k_b$ in their Table I.

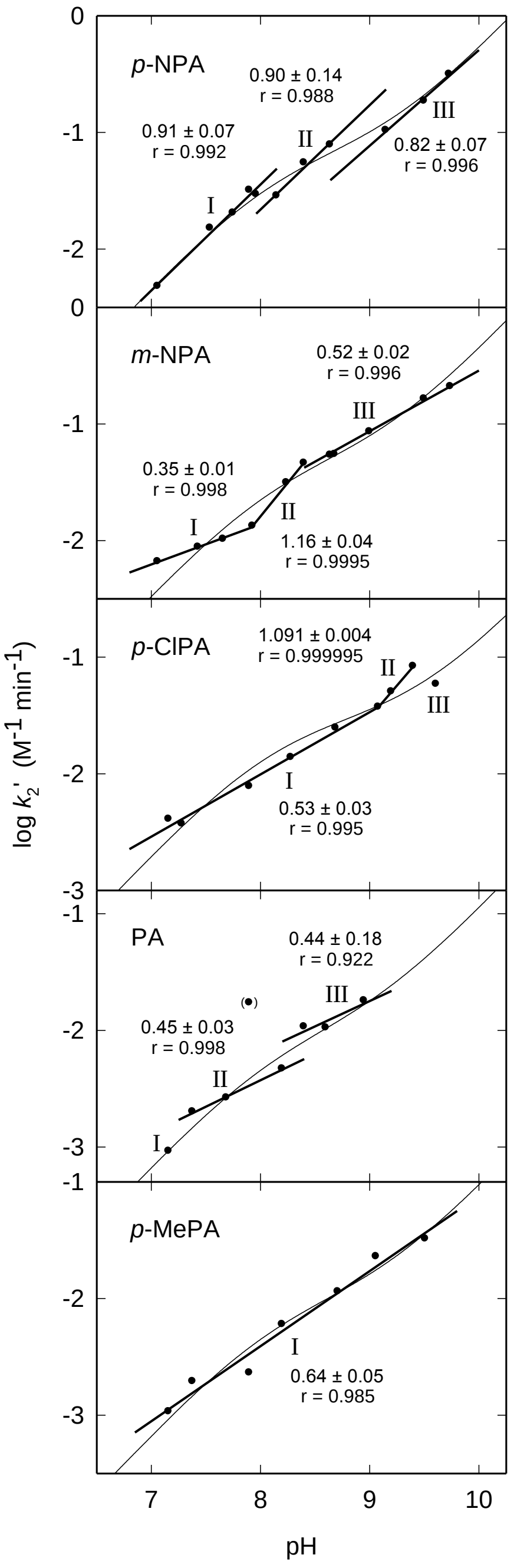

**Fig. 1.** pH profiles for log $k_2$'. Thin lines – authors' representations; thick lines – multiphasic representations. Phase numbers, r values, and slopes ± SE are indicated.

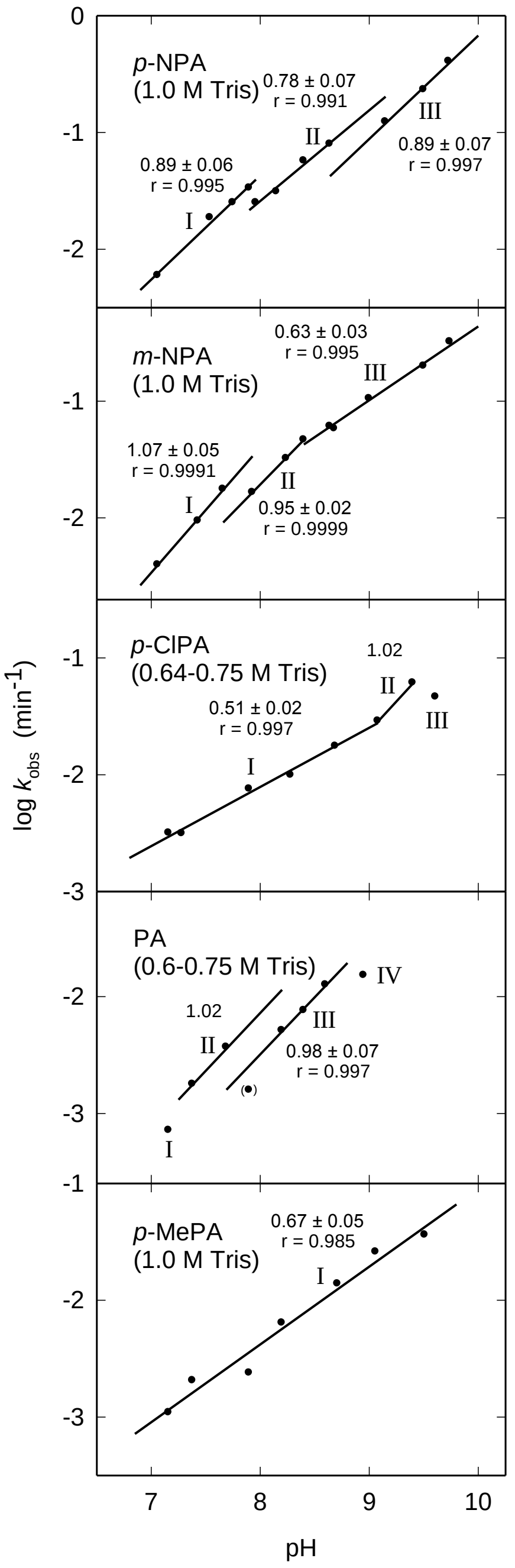

**Fig. 2.** Multiphasic pH profiles for log $k_{obs}$. See also legend to Fig. 1.

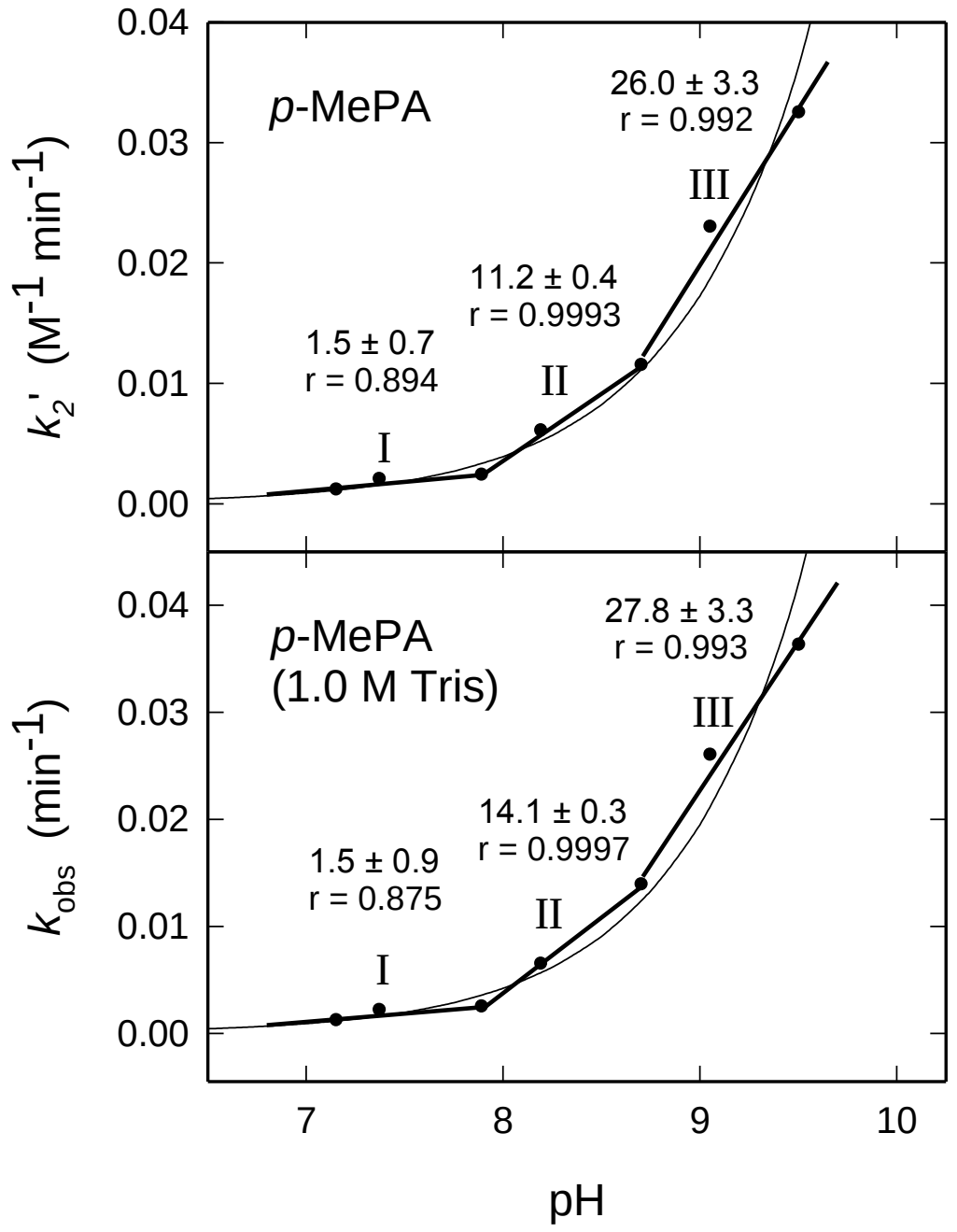

**Fig. 3.** pH profiles for $k_2$’ and $k_{obs}$. Thin curvilinear lines – straight line in plots of log $k_2$’ and log $k_{obs}$ (from Fig. 1); thick lines – multiphasic representations. See also legend to Fig. 1. Slopes multiplied by 1000.

**Table 1**

MSDx1000

| | Authors’ representation | Multiphasic log $k_2$’ vs. pH | Multiphasic $k_2$’ vs. pH |
|---|---|---|---|
| *p*-NPA | 24.6 | 10.5 | 12.1 |
| *m*-NPA | 67.6 | 1.7 | 4.8 |
| *p*-ClPA | 76.8 | 9.4 | 13.8 |
| PA | 46.6 | 15.5 | 22.2 |
| *p*-MePA | 114.8 | 64.6 | 28.1 |

**MSD (mean squares of deviates):** Sum of $((v_{obs} - v_{pred})^2/v_{obs}^2)$/df.
**df for authors’ representation:** Number of points - 2 (for the adjustable parameters $k_a$ and $k_b$).
**df for multiphasic representation:** Sum of df for each phase.

### ***p*-NPA** (*p*-nitrophenyl acetate)

The pH profile for log $k_2$' for this ester can be well represented as triphasic (thick lines in Fig. 1), with the transitions in the form of jumps between pH 7.96 and 8.15, and between pH 8.64 and 9.15. The lines are parallel within experimental error. The fit is markedly better than the fit of the authors' representation (Table 1). The pH profile can be represented as triphasic also in a plot of $k_2$' (not shown). However, lower r values and the somewhat poorer fit (Table 1) indicate that this representation is incorrect.

pH profiles for log $k_{obs}$ are very similar to those for log $k_2$'. Thus, in the presence of the highest TRIS concentration (1.0 M), the profile can, as shown in Fig. 2, also be represented as triphasic, with the transitions between pH 7.90 and 7.96 (jump), and between pH 8.64 and 9.15 (jump). Except for a slight difference in the transition between phases I and II, the patterns for the pH profiles for *p*-NPA in Figs 1 and 2 are identical. The pattern for the pH profile for log $k_{obs}$ in the presence of 0.67-0.75 M TRIS is identical to that in the presence of 1.0 M TRIS (not shown). In the presence of 0.4-0.5 M TRIS the pattern is also triphasic, with the first transition in the form of a jump between pH 7.90 and 7.96, whereas the second transition is at pH 9.15 (not shown).

The point at pH 7.50 in the authors' Fig. 2 appears to be in error, it should have been at pH 7.75. (Their Fig. 3 is correct.)

### ***m*-NPA** (*m*-nitrophenyl acetate)

The pH profile for log $k_2$' can be represented, very precisely, as triphasic (Fig. 1), with the transitions at pH 7.93 and 8.40. The fit is very much better than the fit of the authors' representation (Table 1). The pH profile can be represented, quite precisely, as triphasic also in a plot of $k_2$' (not shown). However, as also for *p*-NPA, the mostly lower r values and the poorer fit (Table 1) indicate that this representation is incorrect.

At the highest TRIS concentration, the pH profile for log $k_{obs}$ can also be well represented as triphasic, with the transitions between pH 7.66 and 7.93 (jump) and at pH 8.40 (Fig. 2). At 0.6-0.75 M TRIS, the pH profile can also be well represented as triphasic, with the transitions in the form of jumps between pH 7.43 and 7.66, and between 8.40 and 8.64 (not shown). At 0.4-0.5 M TRIS, the pH profile can be similarly represented as triphasic, with jumps between pH 7.43 and 7.66, and between pH 8.64 and 8.68 (not shown).

### ***p*-ClPA** (*p*-chlorophenyl acetate)

The pH profile for log $k_2$' can be well represented as triphasic (Fig. 1), with the transitions at pH 9.08, and between pH 9.40 and 9.61. The fit is much better than the fit of the authors' representation (Table 1). The pH profile for $k_2$' can be well represented as tetraphasic (not shown). However, the fit is somewhat poorer (Table 1) and the pattern is more complex (one more phase), indicating that this representation is incorrect.

At 0.9-1.0 M TRIS, the pH profile for log $k_{obs}$ appears to be monophasic (not shown), but there is no point for pH 9.40. At 0.64-0.75 M TRIS, the pH profile can be represented as triphasic, with the transitions at pH 9.08, and between pH 9.40 and 9.61 (Fig. 2), i.e. as in the profile for log $k_2$'. At 0.5 M TRIS, the pH profile can be similarly represented as triphasic, but with the first transition at pH 9.20 (not shown). At 0.25-0.36 M TRIS, the pH profile is tetraphasic (not shown).

There appears to be an error in the authors' Fig. 2. At the highest pH there should be a decrease in $k_2$'.

### **PA** (phenyl acetate)

The pH profile for log $k_2$' can be quite well represented as triphasic (Fig. 1), with the transitions between pH 7.16 and 7.38, and between 8.28 and 8.69 (jump). Lines II and III are precisely parallel. The fit is much better than the fit of the authors' representation (Table 1). The pH profile for $k_2$' can be represented as biphasic, i.e. as less complex than the profile for log $k_2$', but the fit is somewhat poorer (Table 1).

At 1.0 M TRIS, the pH profile for log $k_{obs}$ has only five points, but can be represented as triphasic (not shown). At 0.6-0.75 M TRIS, the pH profile can be represented as triphasic (Fig. 2), with the transitions between pH 7.16 and 7.38, between pH 7.16 and 7.38, between pH 7.69 and 8.20 (jump), and between 8.60 and 8.95. As also found for log $k_2$', lines II and III are parallel, indicating that the profiles have been resolved correctly. At 0.4-0.5 M TRIS, the pH profile can be represented as triphasic, with the transitions between pH 7.16 and 7.38, and as a jump between pH 7.69 and 8.20 (not shown).

A point for pH 7.90 is not shown in the authors' Fig. 2, apparently correctly so. The point (shown in parenthesis in the present plots) is far off the position expected from the other points and has also been omitted from the present calculations.

### ***p*-MePA** (*p*-methylphenyl acetate)

In contrast to the other phenyl esters, the pH profile for log $k_2$' for *p*-MePA cannot be represented as multiphasic. It can be represented as monophasic, but only poorly so (Fig. 1). The point for pH 7.90 is markedly off the line, and there are also other deviations. The fit is better than for the authors' representation, but is still poor (Table 1). The pH profile for $k_2$' can be well represented as triphasic (Fig. 3), with the transitions at pH 7.90 and 8.71. The goodness-of-fit value is still quite high (Table 1), but this is mainly caused by deviations at the two lowest pH values. The fit to triphasic kinetics for $k_2$' is, clearly, better than the fit to monophasic kinetics for log $k_2$' (thin curvilinear line in Fig. 3).

At 1.0 M TRIS, the fit to monophasic kinetics for log $k_{obs}$ is poor, with the point for pH 7.90 again being markedly off the line (not shown). In contrast, the pH profile for $k_{obs}$ can be well represented as triphasic (Fig. 3), with the transitions again at pH 7.90 and 8.71. As also for $k_2$', the fit to triphasic kinetics for $k_{obs}$ is, clearly, better than the fit to monophasic kinetics for log $k_{obs}$. Very similar results (quite poor fits to monophasic kinetics for log $k_{obs}$, good fits to triphasic kinetics for $k_{obs}$) are also found at TRIS concentrations of 0.7-0.75 M, 0.5 M, and 0.25-0.3 M (plots not shown).

## Discussion

Data for pH dependences of cytochromes (Theorell 1941) and yeast pyruvate decarboxylase (Guo et al. 1998) have recently (Nissen 2015) been shown to be better represented by multiphasic profiles than by conventional curvilinear profiles. Very extensive unpublished reanalyses of other pH data, enzymatic as well as non-enzymatic, have given similar results. The finding of identical or very similar multiphasic patterns for several different data sets for each phenyl ester is taken as unequivocal evidence that the present pH profiles have been resolved correctly. The discontinuities in the form of sharp breaks and jumps appear to be identical to those found in pH profiles for enzymatic processes. Parallel lines, as found in the present profiles, occur furthermore quite frequently in a variety of different pH profiles. Thus, from the findings so far, there appears to be no principal differences between pH profiles for enzymatic and non-enzymatic processes. Importantly, the discontinuous transitions appear not to reflect transitions in the protein since they also occur in non-enzymatic systems.

The finding that the pH profiles for *p*-methylphenyl acetate are multiphasic in plots of $k_2$' and $k_{obs}$ vs. pH, whereas the pH profiles for the other and structurally very similar phenyl esters are best represented as multiphasic in plots of log $k_2$' and log $k_{obs}$ vs. pH, is puzzling. In enzymatic systems, rate constants of various kinds are usually best represented as multiphasic in plots of log of the parameter values vs. pH, but there are also data sets where the parameter values vs. pH give the best fits.

In addition to being discontinuous, multiphasic pH profiles are also characterized by the line for each phase being straight, apparently perfectly so when the data are sufficiently precise and the parameters are represented correctly. The reason for this may possibly be related to why straight lines are to be expected in plots of log $K_m$ vs. log $k_{cat}$ for multiphasic relationships, where any change in E+S, ES or $ES^{\#}$ will lead to proportional changes in log $K_m$ and log $k_{cat}$. Straight lines separated by sharp transitions have been found for concentration dependences for solute uptake by plants (Nissen 1974, 1991, 1996) and, in general, for a wide variety of enzyme, uptake and ligand-binding systems from plants, animals and microorganisms (Nissen and Martín-Nieto 1998). Thus, kinetics taken to be cooperative or due to two or more systems may instead be multiphasic, with plots of log $K_m$ vs. log $k_{cat}$ for the various phases forming a straight line.

## Concluding remarks

The finding that pH profiles are multiphasic rather than curvilinear raises a number of fundamental and difficult questions which should now be addressed. In addition to being detailed and precise, pH data should also be rigorously analyzed so that the profiles can be determined correctly. If this is not done it seems that much important information will remain untapped.